\documentclass[12pt]{article}
\usepackage{graphicx}
\usepackage{xcolor}
\usepackage{indentfirst}
\usepackage[font=small]{caption}
\usepackage[symbol]{footmisc}

\newcommand\pubnumber{NuPhys2018-Maturana}
\newcommand\pubdate{\today}
\renewcommand*{\thefootnote}{\fnsymbol{footnote}}
\def\puc{$^1$Instituto de F{\'i}sica,
Pontificia Universidad Cat{\'o}lica de Chile, Avenida Vicu$\tilde{n}$a Mackenna 4860, Santiago, Chile}
\def\utfsm{$^2$Departamento de F{\'i}sica,
Universidad T{\'e}cnica Federico Santa Mar{\'i}a,
Casilla 110-V, Avda. Espa$\tilde{n}$a 1680, Valpara{\'i}so, Chile}
\def\usa{$^3$Amherst Center for Fundamental Interactions, Department of Physics\\
University of Massachusetts Amherst, MA 01003, USA}
\def\support{\footnote{Speaker}}

\def\Title#1{\begin{center} {\Large #1 } \end{center}}
\def\Author#1{\begin{center}{ \sc #1} \end{center}}
\def\Address#1{\begin{center}{ \it #1} \end{center}}

\newcommand\pubblock{\rightline{\begin{tabular}{l} \pubnumber\\
         \pubdate  \end{tabular}}}
\newenvironment{Abstract}{\begin{quotation}  }{\end{quotation}}
\newenvironment{Presented}{\begin{quotation} \begin{center} 
             PRESENTED AT\end{center}\bigskip 
      \begin{center}\begin{large}}{\end{large}\end{center} \end{quotation}}
\def\Acknowledgements{\bigskip  \bigskip \begin{center} \begin{large}
             \bf ACKNOWLEDGEMENTS \end{large}\end{center}}

\def\beq{\begin{equation}}
\def\eeq#1{\label{#1}\end{equation}}
\def\eeqn{\end{equation}}

\def\beqa{\begin{eqnarray}}
\def\eeqa#1{\label{#1}\end{eqnarray}}
\def\eeqan{\end{eqnarray}}

\let\bar=\overbar

\def\Dslash{\not{\hbox{\kern-4pt $D$}}}
\def\dslash{\not{\hbox{\kern-2pt $\del$}}}

\def\msb{{\bar{\ssstyle M \kern -1pt S}}}

\begin{document}
\begin{titlepage}
\pubblock

\vfill
\Title{Towards a way to distinguish between IHDM and the Scotogenic at CLIC}
\vfill
\Author{ \textbf{Ivania Maturana-{\'A}vila}$^{1}$\support , Marco Aurelio D{\'i}az$^1$, Nicol{\'a}s Rojas$^2$, Sebasti{\'a}n Urrutia-Quiroga$^3$}
\Address{\puc}
\Address{\utfsm}
\Address{\usa}

\vfill

\begin{Abstract}
In this talk, a specific collider signature of the Scotogenic model
is presented. We study this signal in order to compare the
similarities and differences between this model and the Inert Higgs Doublet Model (IHDM) under
the light of the forthcoming Compact Linear Collider (CLIC).
\end{Abstract}
\vfill
\begin{Presented}

NuPhys2018, Prospects in Neutrino Physics

Cavendish Conference Centre, London, UK, December 19--21, 2018

\end{Presented}
\vfill
\end{titlepage}
\def\thefootnote{\fnsymbol{footnote}}
\setcounter{footnote}{0}

\section{Introduction}

\indent The Scotogenic Model~\cite{Ma:2006km} is an appealing candidate to solve some of 
the open questions of the Standard Model (SM). 
The model includes radiatively induced neutrino masses and a WIMP-like (Weakly 
interacting massive particle) Dark Matter (DM) candidate which can be either 
scalar or fermion. In the scalar sector, it includes 
the features contained at the Inert Higgs Doublet Model (IHDM), which have 
been widely studied in the literature~\cite{Diaz:2015pyv}. In this work, we study a 
specific collider signal which is present in both models, the Scotogenic Model 
and the  IHDM. We study the differences and resemblances between both considering 
the production of new states at CLIC through the collision between electrons and 
positrons. 

\section{The Models}
\subsection{Inert Higgs Doublet Model}

\indent IHDM includes a second Higgs doublet that will be named $\eta$. Additionally, 
a $Z_2$ symmetry is added in order stabilize the lightest neutral particle 
charged under this symmetry{\color{red},} thus rendering a suitable scalar DM candidate. 
The new scalar field $\eta$ is odd under the $Z_2$ while the SM particles 
are even. Therefore, no tree level \emph{Flavor Changing Neutral Currents (FCNC)} 
arise since new Yukawa interactions between $\eta$ and the fermions of the 
SM are avoided. The most general renormalizable CP conserving scalar potential 
for IHDM is 
\begin{eqnarray}\label{Eq:ScalarPot}
V&=& m_{1}^{2}\phi^{\dag}\phi+m_{2}^{2}\eta^{\dag}\eta+\lambda_{1}(\phi^{\dag}\phi)^{2}+\lambda_{2}(\eta^{\dag}\eta)^{2}+\lambda_{3}(\phi^{\dag}\phi)(\eta^{\dag}\eta)+\lambda_{4}(\phi^{\dag}\eta)(\eta^{\dag}\phi)\nonumber\\
&+&\frac{\lambda_5}{2}(\phi^{\dag}\eta)^{2}+\frac{\lambda_{5}^{\ast}}{2}(\eta^{\dag}\phi)^{2}.
\end{eqnarray}
In order to keep the potential stable, one needs to establish the following conditions 
\begin{equation}
\lambda_{1},\lambda_{2}>0, \quad\quad\lambda_{3},\lambda_{3}+\lambda_{4}-|\lambda_{5}|> -2\sqrt{\lambda_{1}\lambda_{2}}.
\end{equation}
and since we want to fix the real component of the $\eta$ field as the DM candidate, we add that
\begin{equation}\label{Eq:ConditionsDM}
\lambda_4 +\lambda_5 <0 ,\quad\quad \lambda_5 < 0.
\end{equation}
After computing the mass eigenvalues coming out from the $\eta$ field, we will have
\begin{eqnarray}\label{REq:etamasses}
m_{{\pm}}^2 &=& m_2^2 + \frac{\lambda_3}{2}v^{2},\\
m_{{R}}^2 &=&m_2^2 + \frac{\lambda_3}{2}v^{2}+\left(\frac{\lambda_4 + \lambda_5}{2}\right)v^2,\\
m_{{I}}^2 &=& m_2^2 + \frac{\lambda_3}{2}v^{2}+\left(\frac{\lambda_4 - \lambda_5}{2}\right)v^2.
\end{eqnarray}
\subsection{Scotogenic Model}
\begin{figure}[htb!]
\centering
\includegraphics[height=1.5in]{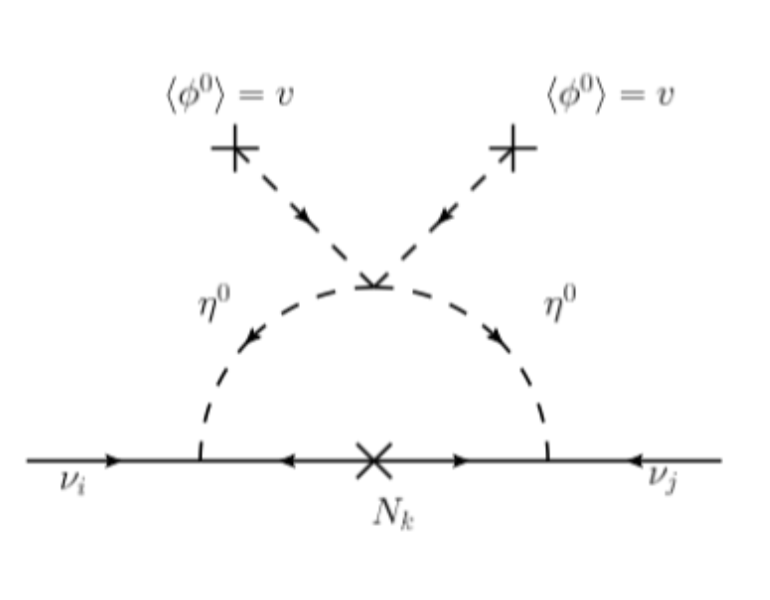}
\caption{Neutrino mass generation at 1-loop.}
\label{fig:figure1}
\end{figure}

\indent The Scotogenic model is considered one of the simplest models containing a DM 
candidate and generating small neutrino masses at 1-loop level. The model has 
a new scalar doublet $\eta$ (with similar properties as the $\eta$ field at IHDM) 
and three singlet fermionic fields $N_i$ $(i=1,2,3)$ all of them odd under $Z_2$. 
The full particle content of the model is listed in the Table~\ref{Table:particlecontent}. 
The neutral components of the $\eta$ field will contribute to neutrino masses at one loop 
(Figure~\ref{fig:figure1}). The resulting formula for the neutrino mass matrix is
\begin{equation}
M_{\nu,ij}=\frac{h_{ik}h_{jk}}{32\pi^{2}}M_{k}\left[\frac{m^{2}_{R}}{m^{2}_{R}-M^{2}_{N_k}}{\log}\left({\frac{m^{2}_{R}}{M^{2}_{N_k}}}\right)-\frac{m^{2}_{I}}{m^{2}_{I}-M^{2}_{N_k}}{\log}\left({\frac{m^{2}_{I}}{M^{2}_{N_k}}}\right)\right].
\end{equation} 
\indent The scalar potential in Scotogenic model is given by the equation~(\ref{Eq:ScalarPot}). 
The new terms at the Lagrangian of the Scotogenic model are a Majorana mass term 
for the $N_{i}$ fields given by $-1/2(\bar{N}_{i}M^{ij}N_{j}^{C})$, and a Yukawa 
coupling $L_{Yukawa} =-h_{ij}\bar{N_i}(i\sigma2\eta^{\ast})^{\dag}L_{j} + h.c$. 
Recall that in both models, we will consider $\eta_R$ as the DM candidate.
It is worth mentioning that in the Scotogenic model the running of 
$m_{2}$ might break the $Z_{2}$ symmetry~\cite{Merle:2016scw} at lower scales of energy, risking the phenomenology of the DM. By studying additional 
considerations to \cite{Merle:2016scw} this problem can be safely avoided.


\begin{table}[htb!]
\centering
\begin{tabular}{| c| c | c | c | c | c | c |}
\hline
& \multicolumn{3}{ c |   }{\quad Standard Model \quad} &  \multicolumn{1}{ c |}{\quad Fermions \quad}  & \multicolumn{1}{ c | }{\quad Scalar \quad}  \\
\hline
           & \quad  $L$   & \quad $e$   & $\phi$   & \quad     $N$    & \quad $\eta$  \\
\hline                                                                            
$SU(2)_L$  & \quad   2    & \quad  1    &    2     & \quad      1    & \quad    2      \\
$Y$        & \quad  -1    & \quad -2    &    1     & \quad       0    & \quad    1        \\
$Z_2$      & \quad  $+$   & \quad $+$   &   $+$    & \quad   $-$   & \quad   $-$    \\
$\ell$        & \quad   1    & \quad  1    &    0     & \quad    0        & \quad    0         \\
\hline
\end{tabular}
\caption{Particle content and quantum numbers for the fields at the Scotogenic model.}
\label{Table:particlecontent}
\end{table}

\section{Results}
Our study subject will be the signature given by the cross 
 section $e^{+}e^{-} \rightarrow \eta^{+}\eta^{-}$, which was 
 computed by using Madgraph~\cite{Alwall:2014hca} and specializing our code for the CLIC experiment~\cite{Roloff:2018dqu}. We will compare that signature in the IHDM and the Scotogenic model. The Feynman diagrams displayed below in Figure~\ref{fig:figure2} show the contributions of each model. 
\begin{figure}[htb]
\centering
\includegraphics[height=1.5in]{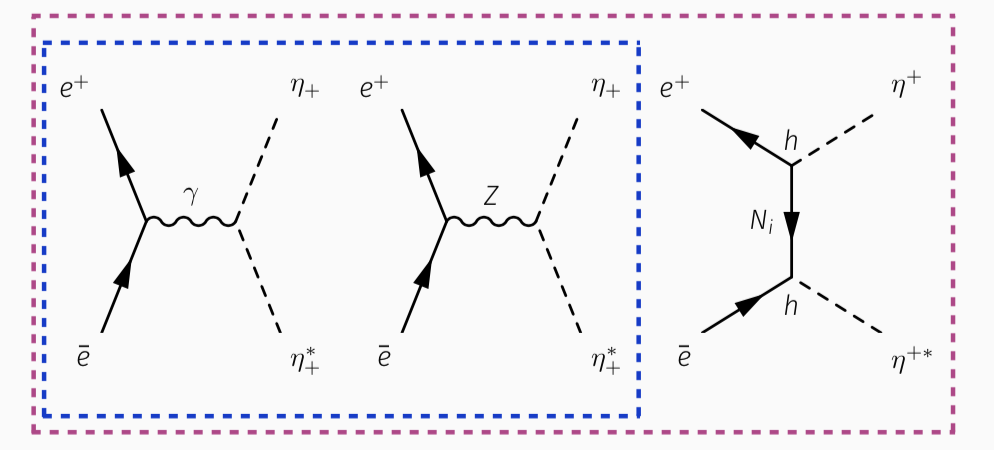}
\caption{Feynman diagrams for IHDM (blue dashed line) and Scotogenic model (Magenta dashed line).}
\label{fig:figure2}
\end{figure}
On the other hand, the plot shown at Figure~\ref{fig:figure3} is based on 
the Benchmark Point given in Table~\ref{Table:BP}. The values displayed 
at that table are in agreement with neutrino physics and preserve $Z_2$ symmetry 
up to scales close to the 8 TeV.
\begin{table}[]
\centering
\begin{tabular}{|cc|}
\hline
$M_{N1} =M_{N2} =[800,1000] $ GeV&   \\
$M_{N3}=1000$ GeV&    \\
$m_2=800$GeV&    \\
$m^{2}_{\nu1}$=[0,0.001] GeV&    \\
$\lambda_{1}=0.26$& \\
$\lambda_{2}=[0,0.5]$&    \\
$\lambda_{3}=0.1$& \\
$\lambda_{4}=-0.1$&    \\
$\lambda_{5}=-10^{-9}$& \\
\hline
\end{tabular}
\caption{Parameter space used for Figure 3.}
\label{Table:BP}
\end{table}
\begin{figure}[htb]
\centering
\includegraphics[height=2.5in]{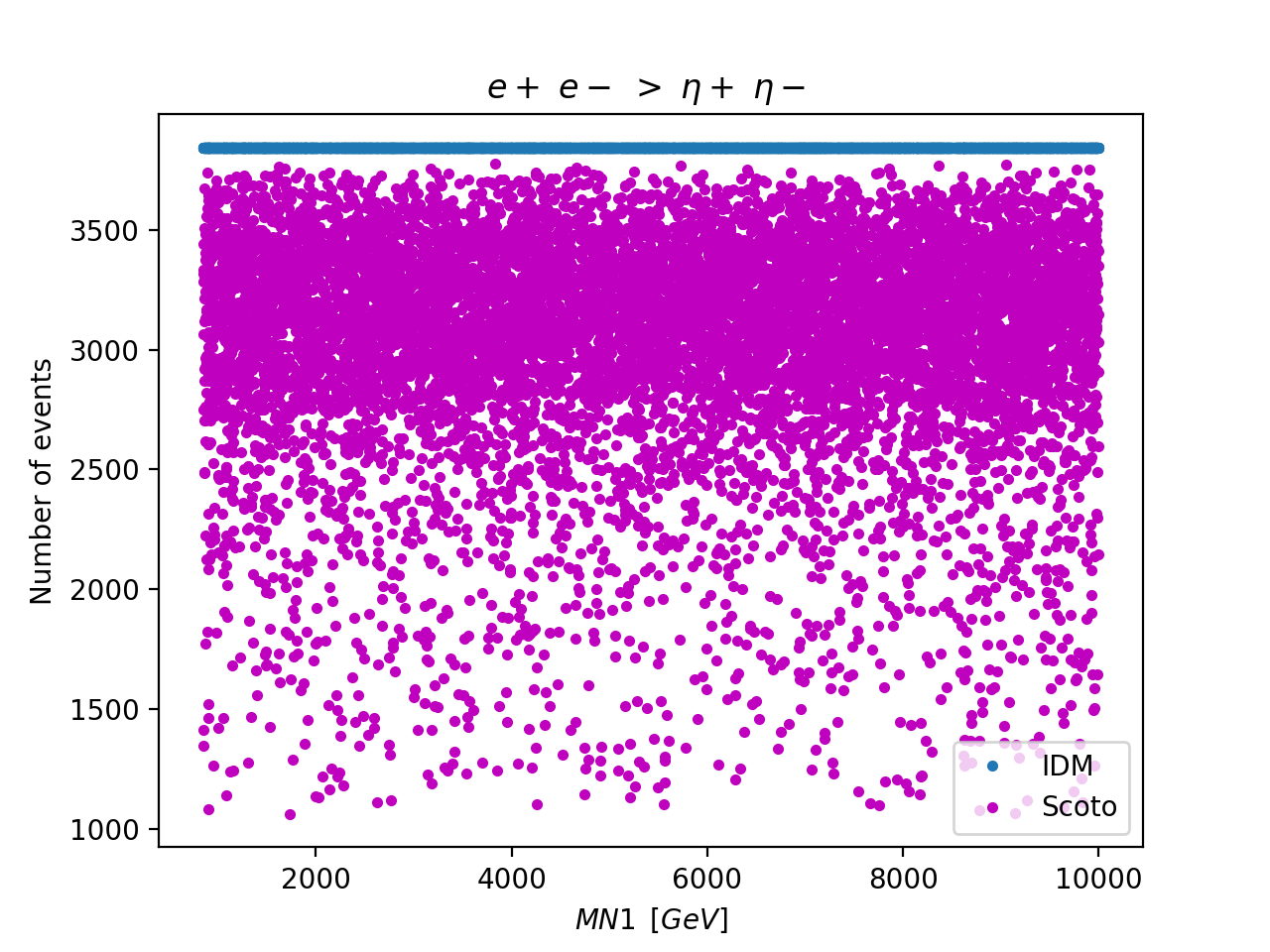}
\caption{Number of events for the two different models: magenta points are the results obtained for the Scotogenic models while blue points are the results obtained for the IHDM }
\label{fig:figure3}
\end{figure}
The number of events resulting from the Scotogenic model (purple points) 
is in general below the IHDM (blue line), which means that process involving 
$N_{i}$ ends up in a destructive interference regarding the contributions 
already present at the IHDM. \\

\indent In conclusion, we have studied the observable $e^+e^-  \rightarrow \eta^+ \eta^- $
under the light of the Scotogenic and IHD (Inert Higgs Doublet) models at CLIC. Constraints 
on the scalar potential have been implemented in order to 
keep both models out from potential destabilization at tree level and radiatively. 
In order to make an extensive comparison to both models, further studies on this 
observable are required. In addition, other signals will be studied
in the future in order to consider constraints coming from DM phenomenology.

\Acknowledgements
IMA was partially supported by CONICYT, Doctorado Nacional (2015) num. 21151255 and Pontificia Universidad Católica de Chile.\\
NR was funded by proyecto FONDECYT Postdoctorado Nacional (2017) num. 3170135.\\
SUQ was partially supported by CONICYT PFCHA/Doctorado Becas Chile/2018 - 72190146.


\begin{thebibliography}{99}


\bibitem{Ma:2006km} 
  E.~Ma,
  Phys.\ Rev.\ D {\bf 73}, 077301 (2006)
  [hep-ph/0601225].
  
\bibitem{Diaz:2015pyv}
  M.~A.~D\'iaz, B.~Koch and S.~Urrutia-Quiroga,
  Adv.\ High Energy Phys.\  {\bf 2016} (2016) 8278375
  [arXiv:1511.04429 [hep-ph]].

\bibitem{Merle:2016scw} 
  A.~Merle, M.~Platscher, N.~Rojas, J.~W.~F.~Valle and A.~Vicente,
  JHEP {\bf 1607}, 013 (2016)
  [arXiv:1603.05685 [hep-ph]].

\bibitem{Roloff:2018dqu} 
  P.~Roloff {\it et al.} [CLIC and CLICdp Collaborations],
  arXiv:1812.07986 [hep-ex].

\bibitem{Alwall:2014hca} 
  J.~Alwall {\it et al.},
  JHEP {\bf 1407}, 079 (2014)
  [arXiv:1405.0301 [hep-ph]].

\end{thebibliography}
\end{document}